\documentclass[12pt,preprint]{aastex}

\def\lsim{\mathrel{\hbox{\rlap{\lower.55ex \hbox {$\sim$}}\kern-.0em\raise.4ex \hbox{$<$}}}} 
\def\gsim{\mathrel{\hbox{\rlap{\lower.55ex \hbox {$\sim$}}\kern-.0em\raise.4ex \hbox{$>$}}}} 
\begin{document}

\def\vla{1}
\def\cit{2}
\def\srl{3}

\def\grb{GRB\,030329}

\title{Accurate Calorimetry of GRB\,030329}

\author{D. A. Frail\altaffilmark{\vla}, A. M.
  Soderberg\altaffilmark{\cit}, S. R. Kulkarni\altaffilmark{\cit}, E.
  Berger\altaffilmark{\cit}, S.  Yost\altaffilmark{\srl}, D. W.
  Fox\altaffilmark{\cit} \&\ F. A.  Harrison\altaffilmark{\srl}}

\altaffiltext{\vla}{National Radio Astronomy Observatory, Socorro, NM
  87801}

\altaffiltext{\cit}{Division of Physics, Mathematics and Astronomy,
  105-24, California Institute of Technology, Pasadena, CA 91125}

\altaffiltext{\srl}{Space Radiation Laboratory 220-47, California
     Institute of Technology, Pasadena, CA\,91125, USA}



\begin{abstract}
  We report late-time observations of the radio afterglow of GRB
  030329.  The light curves show a clear achromatic flattening at 50
  days after the explosion.  We interpret this flattening as resulting
  from the blast wave becoming trans-relativistic.  Modeling of this
  transition enables us to make estimates of the energy content of the
  burst, regardless of the initial jet structure or the distribution
  of initial Lorentz factors of the ejecta. We find, in accordance
  with other events, that \grb\ is well-described by an explosion with
  total energy of a few times 10$^{51}$ erg expanding into a
  circumburst medium with a density of order unity.
\end{abstract}


\section{Introduction}\label{sec:intro}

To fully understand the progenitors of gamma-ray bursts (GRB) and the
workings of the central engines that power them, we must first
understand their energetics.  The wide dispersion in isotropic GRB
energies, once corrected for their jet-like geometry, results in a
narrowly-clustered radiated energy $E_\gamma\simeq 1.3\times 10^{51}$
erg \citep{bfk03,fks+01}. Likewise, the kinetic energy $E_K$ of the
outflow, as derived from the afterglow emission
\citep{wg99,kumar00,fw01}, is clustered around a similar value
\citep{pk02,bkf03}. The close agreement between the values of
$E_\gamma$ and $E_K$ have lead to the prevailing view that GRB
explosions draw from a standard energy reservoir with a total budget
of order a few times 10$^{51}$ erg.

\grb\ appeared to challenge this view.  A distinct break in the
optical and X-ray light curves at $t\sim 0.5$ d \citep{pfk+03,tmg+03}
if interpreted as a jet, gives an opening angle $\theta_j=5^\circ$ and
a radiated energy (corrected for the narrow opening angle)
$E_\gamma=5\times 10^{49}$ erg -- nearly 30 times below the
``standard'' value.  \grb\ is also an order of magnitude
under-energetic as measured by the geometrically-corrected X-ray
luminosity \citep{bkf03}. Other low energy outliers in the E$_\gamma$
distribution \citep{bfk03} and X-ray luminosity distribution do exist.
These energy outliers are important for understanding the diversity of
cosmic explosions. They may represent genuine low energy events, a
long-lived central engine, the viewing of a jet slightly off axis, or
events for which the efficiency of converting the kinetic energy in
the shock to radiation is low.

For \grb\ two different mechanisms have been proposed, both of which
increase increase $E_K$ to about $10^{51}$ erg. To explain the
continued rise of the radio emission over the following week,
\cite{bkp+03} introduced a wide-angle jet ($\theta_j=17^\circ$) with a
lower Lorentz factor than the narrow jet but carrying $\sim 10\times$
more energy. This led \cite{bkp+03} to conclude that GRBs share a
common energy scale {\it only} when the total energy, including that
in mildly relativistic ejecta, is considered. Another variant of the
standard energy hypothesis is to include a long-lived central engine.
This was first proposed to account for an extended bright phase of
afterglow emission of GRB\,021004 \citep{fyk+03}. \cite{gnp03} and
\cite{png04} have argued that such model could also explain the bumps
in the optical light curves \citep{log+04} and the rising radio light
curves (qualitatively at least).

The above discussion shows that there are grounds to question the
simplest hypothesis for GRB explosions. Further progress on the nature
of the central engines will most certainly involve measuring and
modeling the temporal and angular behavior of the energy release.
Nonetheless, the {\it total} energy yield of these explosions is
perhaps the most important bulk parameter of cosmic explosions.
Fortunately, as has been noted in the past \citep{fwk00}, late-time
radio observations offer us precisely this opportunity.  Here, we
report on observations of the radio afterglow of \grb\ on a timescale
of 60 days to 366 days.  Using these observations in conjunction with
the earlier radio data \citep{bkp+03}, we are able to infer a
geometry-independent estimate of the total energy of \grb.

\section{Observations and Results}\label{sec:obs}

\begin{figure}
\plotone{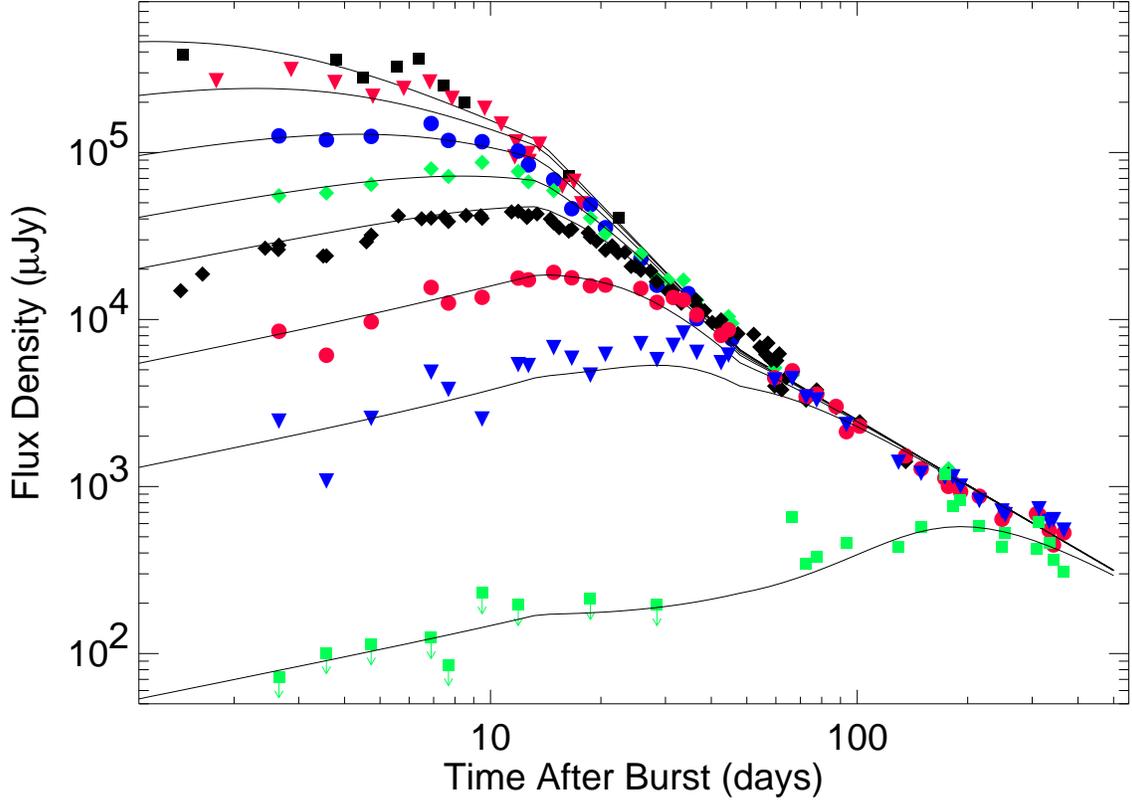}
\caption{Radio light curves of the afterglow of \grb. From bottom to
  top the measurements are were taken at 1.43 GHz (green square), 4.86
  GHz (blue triangle), 8.46 GHz (red circle), 15.3 GHz (black
  diamond), 22.5 GHz (green diamond), 43 GHz (blue circle), 100 GHz
  (red triangle), 250 GHz (black square). The data have been scaled
  relative to the 8.46 GHz measurements by $\nu^\beta$, where $\beta$
  is the spectral slope between 4.86 and 22.5 GHz for $t\gsim 75$ d
  (\S\ref{sec:obs}). The solid lines are the predicted emission from a
  jet expanding into a constant density medium (see
  \S\ref{sec:bbmodel}). The flattening of the light curves beyond
  t$\sim$50 days occurs when the jet expansion becomes
  sub-relativistic.  Although this model captures the gross evolution
  of the observed light curves (the rise to maximum, the initial steep
  decay at high frequencies and the flattening of the optically thin
  light curves at late times), the unprecedented high signal-to-noise
  of this data does, however, show some deficiencies of our model.  In
  addition to the expected fluctuations at low frequencies due to
  interstellar scintillation, there are real variations at higher
  frequencies such as the ``bump'' between 50 and 60 day seen at 15
  GHz. Owing to our simplistic treatment, the model also under predicts
  (or over predicts) the flux density at different transition times
  when the evolution changes abruptly.}
\label{fig:one}
\end{figure}

All observations were made with the Very Large Array
(VLA)\footnotemark\footnotetext{The VLA is operated by the National
  Radio Astronomy Observatory, a facility of the National Science
  Foundation operated under cooperative agreement by Associated
  Universities, Inc.} and reduced following standard practice
\citep[e.g.,][]{fwk00}. A summary of the flux density measurements are
in Table \ref{tab:obs}. In Figure \ref{fig:one} we plot the radio
light curves supplemented with our earlier data from \cite{bkp+03} and
\cite{sfw+03}. 

A significant flattening of the late-time radio light curves at
$t\gsim 50$ d can be seen, especially when compared to the steep
decline measured at early times at higher frequencies (22.5-250 GHz)
where the temporal slope $\alpha\simeq -2$ (with $F\propto t^\alpha$)
at $t>10$ days.  This flattening is achromatic, except at 1.43 GHz
where the light curve was still rising, reaching a maximum near 300 d.
A joint temporal and spectral power-law fit of the form
$F(\nu,t)\propto t^\alpha \nu^\beta$ for $t\gsim 75$ d at 4.86, 8.46,
15 and 22.5 GHz gives $\alpha=-1.20 \pm 0.06$ and $\beta=-0.58\pm
0.08$.  A similar flattening of the X-ray emission is seen by {\it
  XMM-Newton} in which the decay index changed from $\alpha_X=-1.86
\pm 0.06$ to $\alpha_X=-1.40 \pm 0.15$ between 1.24 and 37 days and 37
and 258 days after the burst \citep{tmg+03,tmg+04}. The late-time
X-ray spectral index $\beta_X=1.17\pm 0.04$. The difference in the
radio/X-ray spectral slopes, $\beta-\beta_X=0.59\pm0.09$ is of the
right magnitude ($\Delta\beta=0.5$) if these bands lie on either side
of the synchrotron cooling frequency, {\em i.e.} $\nu_R<\nu_c<\nu_X$.

The simplest explanation for an achromatic flattening of afterglow
light curves at late times is that the blast wave has become
trans-relativistic.  Such a transition occurs when the rest mass
energy swept up by the expanding shock becomes comparable to the
initial kinetic energy of the ejecta.  Given the jet parameters of
\grb, this is predicted to occur roughly at a time t$_{NR}\simeq
t_j/\theta_j^2\simeq 60$ d \citep{fwk00}.

In the non-relativistic regime the temporal index in a constant
density medium (ISM) is $\alpha_{NR}=(21-15p)/10$ for $\nu<\nu_c$ or
$\alpha_{NR}=(4-3p)/2$ for $\nu>\nu_c$; here $p$ is the energy index
of the electrons accelerated into a shock, and it is related to the
spectral slope $p=-2\beta$ for $\nu>\nu_c$, and $p=1-2\beta$ for
$\nu<\nu_c$. In contrast, in the relativistic regime, the flux is
expected to steeply decline, $t^{-p}$ independent of frequency.
Similar expressions exist for a a wind-blown medium (WIND) whose
density varies as the inverse square of the radius:
$\alpha_{NR}=(5-7p)/6$ for $\nu<\nu_c$ or $\alpha_{NR}=(8-7p)/6$ for
$\nu>\nu_c$ \citep{lw00}.

Applying the above expressions for $\nu<\nu_c$ to the radio
measurements of $\alpha$ and $\beta$ gives $p=2.16\pm0.16$,
$\alpha_{NR}=-1.14\pm 0.24$ (ISM) and $\alpha_{NR}=-1.69\pm 0.19$
(WIND). Thus the observed flattening of the radio light curves
($\alpha=-1.20 \pm 0.06$) is consistent with a constant density
medium, but a wind-blown medium predicts a decay that steeper than
expected. A wind-blown medium is rejected at the 2.5$\sigma$ level.
Applying the same expressions to the X-ray measurements where
$\nu>\nu_c$, we derive $p=2.34\pm0.08$, $\alpha_{NR}=-1.51\pm 0.18$
(ISM) and $\alpha_{NR}=-1.40\pm 0.09$ (WIND). By themselves, the X-ray
data is not able to distinguish between models while the radio favors
the ISM model.

Given this good agreement (both in the timescale and the magnitude of
the flattening) with the expectations of a jet expanding in a constant
density medium, we are motivated to consider more detailed models in
the following sections.

\section{Modeling the Broadband Afterglow}\label{sec:bbmodel}

Initially the blast wave is expanding relativistically and thus we can
use the extensive machinery developed for afterglow modeling
\citep[e.g.,][]{wg99,cl00,pk02}. The details of our specific method
are given in \cite{yhsf03}. Briefly, the blast wave is assumed to
accelerate a power-law distribution of electrons
N($\gamma_e)\propto\gamma_e^{-p}$ with energy index $p$ above some
minimum energy $\gamma_m$. The emitted spectrum (synchrotron and
possibly inverse Compton) at any time $t$ can be described by the peak
flux and several break frequencies whose evolution depends on the
shock dynamics.

Prior to the jet break time $t_j$ the evolution is identical to a
spherically symmetric shock \citep{spn98}, while for $t_j<t<t_{NR}$
the evolution is described by a jet expanding in a constant density
medium \citep{sph99}. After $t_{NR}$ the evolution of the break
frequencies is governed by Sedov-Taylor dynamics
\citep[e.g.,][]{fwk00} and discussed in the next section.  To allow a
smooth transition of the light curves, the normalization of the
emission spectrum is determined by the equations for the jet model at
$t=t_{NR}$.

A total of six physical parameters were fitted for; the isotropic
equivalent energy $E_{52}$ of the jet, its half opening angle
$\theta_j$, the circumburst density $n$, the electron energy index
$p$, and the fraction of the shock energy in electrons $\epsilon_e$
and in magnetic fields $\epsilon_B$. We used only the radio and X-ray
data since the optical data are dominated by the supernova 2003dh at
late times.

We show the results of our fit in Figure \ref{fig:one} and summarize
the model parameters in Table \ref{tab:bestfits}.  Also in Table
\ref{tab:bestfits} is the best-fit jet model from \cite{bkp+03}.  The
calculation of $\theta_j$ from \cite{bkp+03} was originally derived
using the isotropic gamma-ray energy and some {\it assumed} radiative
efficiency. To be internally consistent we have re-computed $\theta_j$
in Table \ref{tab:bestfits} directly from the $E_{52}$ and $n$ using
the formalism of \cite{sph99}.

\section{Modeling the Afterglow at Late Times}\label{sec:nrmodel}

As noted above, after $t_{\rm NR}$ the blast wave becomes
non-relativistic.  On about the same timescale the side-ways expansion
of the jet becomes important and eventually the blast wave becomes
spherical.  Modeling the afterglow emission at late stages thus offer
the distinct advantage of the inferred parameters being independent of
the relative orientation of the axis of the jet and the line of sight
to the observer.

We follow the methodology of \citet{fwk00} (Appendix A) and
\cite{ber04}. Essentially the primary observable is the synchrotron
self-absorbed radio spectrum with the peak  flux density ($f_{\nu}$),
the optical depth ($\tau_{\nu}$) and the characteristic synchrotron
frequency ($\nu_{m}$). These parameters, in turn, are determined by the
values of four physical parameters at $t=t_{\rm NR}$: the magnetic
field ($B_{\rm NR}$), the shock radius ($r_{\rm NR}$), the minimum
Lorentz factor of the electrons ($\gamma_{m,\rm NR}$) and the number
density of the radiating electrons ($n_{e,\rm NR}$).

The evolution of the self-absorbed spectrum is governed by the
dynamics of the blast wave. In particular, at late times the blast
wave evolution is well described by the Sedov-von Neumann-Taylor (SNT)
solution \citep{snt}: $r=r_{\rm NR}(t/t_{\rm NR})^{2/5}$; speed (in
units of speed of light, $c$) $\beta=\beta_{\rm NR}(t/t_{\rm
  NR})^{-3/5}$; $B=B_{\rm NR}(t/t_{\rm NR})^{-3/5}$ and
$\gamma_m=\gamma_{m,\rm NR}(t/t_{NR})^{-6/5}$. We drop the NR
subscript hereafter for clarity.

There are four unknown parameters but usually only three observables
(but see \S\ref{sec:cool}).  Thus we must introduce additional
constraints to fully specify a solution.  We consider three possible
constraints and these are discussed in turn.

\subsection{Equipartition}\label{sec:equi}

The standard approach in radio astronomy is to appeal to equipartition
for an additional constraint. This means equating the fraction of
energy in relativistic electrons, $E_e=[(p-1)/(p-2)]\times n_e
\gamma_m m_e c^2 V$ to the energy in magnetic fields, $E_B=B^2 V/8\pi$
or $\epsilon_e=\epsilon_B$.  Here, $V$ is the volume within the shell
of emitting electrons, $V=4\pi r^3/\eta~\rm cm^{3}$, where $r/\eta$ is
the thickness of the shell and we use the standard (thin shell)
assumption that $\eta\approx 10$. The density $n= n_e\times 3/\eta$
where $3/\eta$ is the shock compression factor.

Under the assumption of equipartition we find the following best-fit
solution for rest frame physical parameters: $r_{ep}\approx 3.5\times
10^{17}~\nu_{m,9}^{-0.0059}$ cm, implying $B_{ep}\approx
0.18~\nu_{m,9}^{-0.024}$ G, $\gamma_{m,ep}\approx
42~\nu_{m,9}^{0.51}$, $n_{e,ep}\approx 5.9~\nu_{m,9}^{-0.56}~\rm
cm^{-3}$ and
$\epsilon_{e,ep}~\nu_{m,9}^{0.52}=\epsilon_{B,ep}~\nu_{m,9}^{0.52}\approx
0.08$ for values of $p\approx 2.2$ and $t_{\rm NR}\approx 50$ days. 

As a consistency check, we note that applying Equation 13 of
\citep{fwk00} gives $\beta_{ep}\approx 1.25~\nu_{m,9}^{-0.0059}$ and
thus we confirm that it is appropriate to apply the SNT solution at
$t_{\rm NR}$. 

Because $\nu_m$ drops below our lowest frequency just before $t_{\rm
  NR}$, the model parameters are given in terms of $\nu_{m,9}$ which
is normalized to 1 GHz at $t_{\rm NR}$ for $p=2.2$, in accordance with
our relativistic modeling (\S\ref{sec:bbmodel}).  Using the model
constraints above we derive $E_{\rm min}=E_{e,ep}+E_{B,ep}\approx
1.3\times 10^{50}~\nu_{m,9}^{-0.065}$ erg. The total energy in the
ejecta (including shocked protons), derived from the Sedov-Taylor
energy equation is $E_{\rm ST,ep}\approx 8.5\times
10^{50}~\nu_{m,9}^{-0.59}$ erg.

\subsection{Consistency with the Sedov-Taylor Solution}

Instead of adopting the equipartition assumption in \S\ref{sec:equi}, we
can use the more general constraint that the energy in the electrons
and magnetic field does not exceed the thermal energy \citep{fwk00},
$E_{e}+E_{B} \le E_{\rm ST}/2$.  This constraint is a logical
consequence of our assumption that the blast wave can be described by
a Sedov-Taylor solution at late times.

\begin{figure}
\plotone{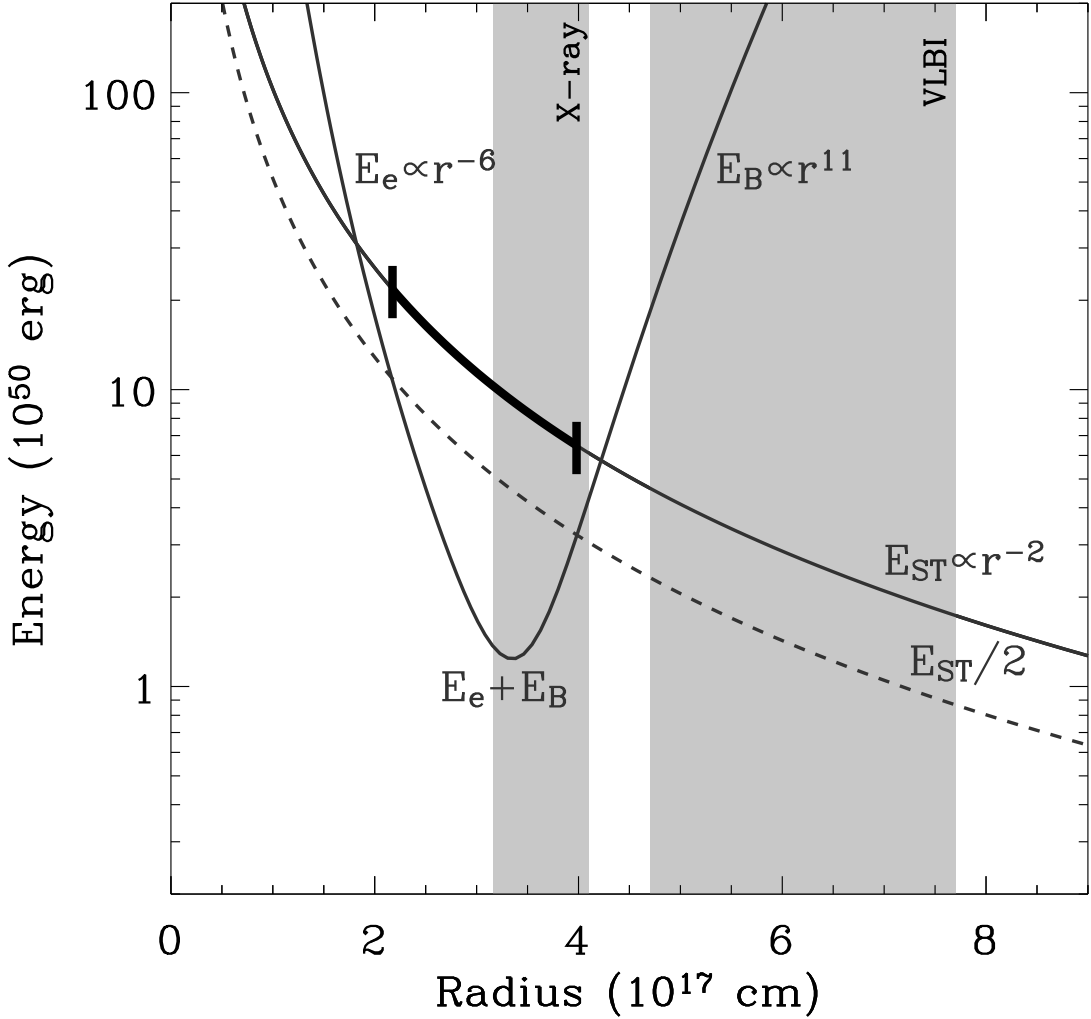}
\caption{Dependence of energy on radius at $t=t_{\rm NR}\approx 50$
  days as described in Section~\ref{sec:nrmodel}.  The sum of the
  energy in electrons and magnetic fields, $E_e+E_B$, is minimized
  near the equipartition radius.  The Sedov-Taylor energy scales with
  the unknown radius as $E_{\rm ST}\propto r^{-2}$.  The permitted
  range of kinetic energy values are defined where $(E_e+E_B) \le
  E_{\rm ST}/2$ and are marked by the bounded solid line.  Here we are
  adopting $\nu_{m,\rm NR}=$1 GHz.  The radius of the shock, as
  measured from VLBI observations of \cite{tfb+04} at $t\approx 83.3$
  days (and scaled down to $t\approx 50$ days as $r\propto t^{0.4}$)
  is shown by the vertical shaded band; it is a factor of $\sim 2$
  times larger (1.8$\sigma$ significance) than that predicted by our
  NR modeling.  Including the additional constraint of the X-ray
  observations \citep{tmg+04}, we are able to constrain the radius to
  $3.6\pm 0.5\times 10^{17}$ cm (vertical shaded band) -- nearly equal
  to the equipartition value.  This implies a slightly lower energy of
  $E_{\rm ST}\approx 7.8^{+2.2}_{-1.6}\times 10^{50}$ erg.}
\label{fig:EnergyRadius}
\end{figure}

For radii above (below) $r_{ep}$, the summed energy rises steeply as
$r^{11}$ ($r^{-6}$) when $E_{B} > E_{e}$ ($E_{B} < E_{e}$).  With this
method we can probe the parameter space {\it away from} equipartition
and derive a range of allowed values for $E_{ST}$.
Figure~\ref{fig:EnergyRadius} shows the dependence of $E_{\rm ST}$ and
$(E_{e}+E_{B})$ on radius. As can be seen from the Figure, the
permitted values of the Sedov-Taylor energy are $E_{\rm ST}\approx
0.64 - 2.2\times 10^{51}$ erg, corresponding to radius values of
$r\approx 2.2 - 4.0\times 10^{17}$ cm.  This range of radii implies
the following ranges for the physical parameters: $B\approx 0.027 -
0.30$ G, $n_{e}\approx 2.3 - 160~\rm cm^{-3}$, $\beta\approx 0.78 -
1.4$, $\epsilon_{e}\approx 4.4\times 10^{-2} - 0.49$,
$\epsilon_{B}\approx 1.7\times 10^{-4} - 0.45$.

\subsection{The Fourth Observable - Cooling Frequency}\label{sec:cool}

Late-time X-ray observations \citep{tmg+04} as in \S\ref{sec:bbmodel}
allow us to constrain the cooling frequency $\nu_c$, and hence
determine the radius of the blastwave. In particular, X-ray
observations allow us to model the evolution of the cooling frequency,
$\nu_c\propto B^{-3} t^{-2}$ \citep{fwk00}. For the same value of
$\nu_{m,9}$ used above: $r\approx 3.6\times 10^{17}$ cm, $B\approx
0.21$ G, $\gamma_{m}\approx 39$, $n_{e}\approx 4.3~\rm cm^{-3}$ and
$\epsilon_{e}\approx 0.064$, $\epsilon_{B}\approx 0.13$ (Table
\ref{tab:bestfits}). The total energy in the ejecta $E_{\rm
  ST}\approx{7.8}^{+2.2}_{-1.6}\times 10^{50}$ erg.  The uncertainty
in the energy is dominated by the flux density error of the X-ray
observations at $t\sim 60$ days.

\section{Angular Size of the Afterglow}

Comparison of our derived size with the observed size offers the
opportunity to directly verify models. Normally such a comparison is
inaccurate since the expected angular sizes are small and until
recently, they were inferred from interstellar scintillation
observations \citep[e.g.,][]{fkn+97}. However, \cite{tfb+04} have
directly determined the size of the fireball for \grb, measuring an
angular diameter $\theta=172\pm43$ $\mu$as at $t=83.3$ d. 

Using Sedov-Taylor expansion ({\em i.e.,} $r\propto t^{2/5}$) to scale
to $t_{NR}$=50 d, and assuming a Lambda cosmology ($H_0 =
71$~km/s/Mpc, $\Omega_M = 0.27$, $\Omega_\Lambda=0.73$), the
angular-diameter distance of \grb\ at $z=0.1685$ is $d_A=589\,$Mpc and
its radius $r=(6.2\pm 1.5)\times 10^{17}$ cm. Thus it appears that the
VLBA measurement is at odds with the radius estimates deduced above,
albeit at only 2$\sigma$ significance (see
Figure~\ref{fig:EnergyRadius}). To better understand this difference
(and to extract more robust estimates of $r$ and $E_{ST}/n$), it would
be useful to jointly fit the evolution of the apparent size of \grb\ 
\citep{onp04} together with the evolution of the afterglow emission.

\section{Discussion and Conclusions}

Continued monitoring of the afterglow of \grb\ has revealed an
achromatic flattening (\S\ref{sec:obs}) of the light curves whose
magnitude and timescale are consistent with a jet-like outflow which
is undergoing a dynamical transition to trans-relativistic expansion.
We have modeled the entire dataset (1-366 d) and derive a total
kinetic energy of $9.0\times 10^{50}$ erg (\S\ref{sec:bbmodel}). This
estimate agrees with a previous estimate of $6.7\times 10^{50}$ erg,
based on observations from 1-64 d \citep{bkp+03}. Thus our analysis is
consistent with earlier conclusions that \grb\ was not an
under-energetic burst \citep{bkp+03,gnp03}.

We also attempted to derive an independent estimate of the energy
using only those observations beyond $t>t_{NR}$. The Sedov-Taylor
analysis (\S\ref{sec:nrmodel}) traces {\it all} ejecta regardless of
the initial jet structure or the distribution of initial Lorentz
factors of the ejecta. This accurate calorimetric method has now been
applied to \grb, GRB\,970508 and GRB\,980703 \citep{fwk00,ber04}.  In
this particular case, the value of this approach is limited by the
uncertainty in the exact value of the synchrotron peak frequency
$\nu_m$. Using the X-ray data as an added constraint
(\S\ref{sec:cool}), and adopting $\nu_m$=1 GHz at $t_{NR}$ we derive
an energy of $7.8\times 10^{50}$ erg.

Strictly speaking the kinetic energies given above are upper limits
since they are derived at a fixed time ($t_{NR}$). The fireball
undergoes radiative losses especially prior to the time when the
evolution is in the fast cooling regime \citep{spn98}.  We estimate
from our full model in \S\ref{sec:bbmodel} that this fast-to-slow
cooling transition occurs at $t=0.03$ d. At this time the total
isotropic energy $E_{52}=2.0\times 10^{52}$ erg and the
geometrically-corrected energy E$_{K}=2.2\times 10^{51}$ erg.  The
energy (isotropic or otherwise) drops to 42\% of this value at $t_j$
and 37\% at $t_{NR}$.

With further observations of \grb\ we have an opportunity to look for
departures in the standard blastwave model. At present, the physical
parameters of the blastwave derived from only the first two months
\citep[Table \ref{tab:bestfits} and][]{bkp+03} provide a remarkably
good prediction of its subsequent behavior for up to a year later.
Deviations may be expected to occur at late times because of an
injection of additional energy by slower moving ejecta, an evolution
in the microphysical parameters of the shock ($p$, $\epsilon_e$,
$\epsilon_B$) with time, or a change in the density structure of
circumburst medium.

The current estimates of the energy and other physical parameters of
\grb\ are accurate at best to a factor of two. While a more
quantitative analysis \citep[e.g.,][]{yhsf03} is needed, we can
already rule out order of magnitude changes in E$_{K}$, $\epsilon_e$
and $\epsilon_B$. Identifying changes in the density structure is more
difficult because between 1 day and 1 year the blast wave samples only
a small range of radii in its rest frame ($\sim$0.1 to 0.4 pc).
However, at its present rate of decay the afterglow of \grb\ will be
visible at centimeter wavelengths for the next decade. Thus the
prospects for tracing the continued evolution of \grb\ are excellent.

\acknowledgments

DAF wishes to thank Pawan Kumar for his hospitality at the University
of Texas, Austin during which time this paper was begun. We
acknowledge NSF and NASA grants for support.


\begin{deluxetable}{lrrrrrc}
\tablecolumns{7}
\tablewidth{0pt}
\tablecaption{Observational Summary\label{tab:obs}}
\tablehead{
\colhead{Date}&
\colhead{$\Delta t$}&
\colhead{$F_{1.43}$}&
\colhead{$F_{4.86}$}&
\colhead{$F_{8.46}$}&
\colhead{$F_{15.0}$}&
\colhead{$F_{22.5}$}\\
\colhead{(UT)}&
\colhead{(days)}&
\colhead{($\mu$Jy)}&
\colhead{($\mu$Jy)}&
\colhead{($\mu$Jy)}&
\colhead{($\mu$Jy)}&
\colhead{($\mu$Jy)}\\
\colhead{(1)}&
\colhead{(2)}&
\colhead{(3)}&
\colhead{(4)}&\colhead{(5)}&
\colhead{(6)}&
\colhead{(7)}}
\startdata
2003 Jun. 10.00 & 72.52 & 1009$\pm$116 & 4808$\pm$67 & 3454$\pm$69 & 2328$\pm$207 & \nodata \\
2003 Jun. 15.00 & 77.52 & 1111$\pm$84  & 4613$\pm$66 & 3572$\pm$54 & 2680$\pm$200 & \nodata \\
2003 Jun. 25.01 & 87.53 & \nodata      & \nodata     & 3016$\pm$31 & 2100$\pm$165 & 1650$\pm$65 \\
2003 Jun. 30.94 & 93.46 & 1360$\pm$40  & 3290$\pm$57 & 2130$\pm$55 & \nodata & \nodata \\
2003 Jul. 09.02 & 101.54& \nodata      & \nodata     & 2299$\pm$65 & 1729$\pm$182 & 1290$\pm$65 \\
2003 Jul. 27.88 & 120.40& \nodata      & \nodata     & 1190$\pm$35 & \nodata & \nodata \\
2003 Aug. 06.06 & 129.58& 1276$\pm$69  & 1955$\pm$62 & \nodata     & \nodata & \nodata\\
2003 Aug. 11.96 & 135.48& \nodata      & \nodata     & 1525$\pm$56 & 997$\pm$171  & 794$\pm$119\\
2003 Aug. 25.89 & 149.41& 1677$\pm$66  & 1677$\pm$55 & 1277$\pm$51 & \nodata & \nodata \\
2003 Sep. 18.79 & 173.31& 3470$\pm$65  & 1590$\pm$52 & 1122$\pm$45 & \nodata & \nodata \\
2003 Sep. 22.74 & 177.26& \nodata      & \nodata     & 1002$\pm$91 & 436$\pm$200  & 374$\pm$166 \\
2003 Sep. 27.75 & 182.27& 2240$\pm$66  & 1600$\pm$46 & 1050$\pm$38 & \nodata & \nodata \\
2003 Oct. 06.73 & 191.25& 2450$\pm$58  & 1407$\pm$44 & 931$\pm$41  & \nodata & \nodata \\
2003 Oct. 30.72 & 215.24& 1709$\pm$52  & 1154$\pm$43 & 874$\pm$44  & \nodata & \nodata \\
2003 Dec. 02.58 & 248.10& 1276$\pm$40  & 999$\pm$32  & 637$\pm$33  & \nodata & \nodata \\
2003 Dec. 07.58 & 253.10& 1541$\pm$41  & 949$\pm$32  & 696$\pm$31  & \nodata & \nodata \\
2004 Jan. 31.53 & 308.05& 1240$\pm$54  & \nodata     & 686$\pm$38  & \nodata & \nodata \\
2004 Feb. 05.36 & 312.88& 1800$\pm$53  & 1029$\pm$26 & 675$\pm$24  & \nodata & \nodata \\
2004 Feb. 26.27 & 333.79& 1360$\pm$96  & 865$\pm$33  & 547$\pm$27  & \nodata & \nodata \\
2004 Mar. 06.33 & 342.85& 1075$\pm$124 & 886$\pm$45  & 447$\pm$33  & \nodata & \nodata \\
2004 Mar. 29.41 & 365.93&  904$\pm$102 & 767$\pm$34  & 528$\pm$29  & \nodata & \nodata \\
\enddata

\tablenotetext{*}{\small Table columns are (1) Epoch of observation; (2) time
  in days from burst, (3) flux density at 1.43 GHz, (4) flux density
  at 4.86 GHz, (5) flux density at 8.46 GHz, (6) flux density at 15
  GHz, and (7) flux density at 22.5 GHz.}

\tablecomments{To maximize sensitivity, the full VLA bandwidth (100
  MHz) was recorded in each of two hands of circular polarization.
  The amplitude scale was calibrated with observations of J1331+303
  (3C\,286), while phase calibration was achieved with frequent
  observations of J1051+2119 at all frequencies except for 1.43 GHz
  where J1016+2037 was used instead.}

\end{deluxetable}

\clearpage
\newpage
\begin{deluxetable}{ccccccccccccc}
\tablecolumns{13}
\tablewidth{0pc}
\tablecaption{Best Fit Parameters For Different Models\label{tab:bestfits}}
\tablehead{
\colhead{} &
\colhead{Range} &
\colhead{} &
\colhead{$t_{j}$} &
\colhead{$t_{NR}$} &
\colhead{$E_{K}$\tablenotemark{c}} &
\colhead{$\theta_{j}$} &
\colhead{$n$} &
\colhead{} &
\colhead{} &
\colhead{$\epsilon_{B}$} \\
\colhead{Model\tablenotemark{a}} &
\colhead{(d)} &
\colhead{$\chi^2$/dof\tablenotemark{b}} &
\colhead{(d)} &
\colhead{(d)} &
\colhead{(erg)} &
\colhead{(rad)}&
\colhead{(cm$^{-3}$)} &
\colhead{$p$} &
\colhead{$\epsilon_e$} &
\colhead{(\%)}
}
\startdata
Rel. & 1-64  & 5133/164 & 10 & N/A & 0.67 & 0.45 & 3.0 & 2.2  & 0.19 & 4.2 \\
Full & 1-366 & 3245/241 & 14 & 48 & 0.90 & 0.47 & 2.2 & 2.2 & 0.17 & 7.4 \\
Non-Rel & 50-366 & 5313/75 & N/A & 50 & 0.78 & N/A & 1.3 & 2.2  & 0.06 & 13 \\
\enddata

\tablenotetext{a}{From top to bottom, the parameters are given for the
  relativistic model by \cite{bkp+03}, the full evolution model from
  \S\ref{sec:bbmodel}, and the non-relativistic model from
  \S\ref{sec:cool}.}

\tablenotetext{b}{The large $\chi^2$ for these fits is affected by
  intensity variations induced by density inhomogeneities in the
  ionized interstellar medium of our galaxy \citep{goo97}.
  Diffractive scintillation dominates at early times, while the
  fluctuations in the late-time 1.43 GHz light curve are likely due to
  large scale refractive effects.}

\tablenotetext{c}{The blastwave kinetic energy in units of $10^{51}$
  ergs. The first two models have been corrected for collimation by
  multiplying the isotropic equivalent energy by $(1-\cos\theta_j)$.
  The energies for the first two models are calculated at $t_j$, while
  the last model is the Sedov-Taylor energy calculated at $t_{NR}$.}

\end{deluxetable}

\clearpage

\clearpage

\end{document}